# Performance of MEMS-based visible-light adaptive optics at Lick Observatory: Closed- and open-loop control

Katie Morzinski[*a,b], Luke C. Johnson[a,b], Donald T. Gavel[a,b], Bryant Grigsby[c], Daren Dillon[a,b], Marc Reinig[a,b], and Bruce A. Macintosh[a,d]

[a] Center for Adaptive Optics, University of California, 1156 High St., Santa Cruz, CA 95064;
[b] Laboratory for Adaptive Optics, Univ. of California, 1156 High St., Santa Cruz, CA 95064;
[c] Univ. of California Observatories/Lick Observatory, P.O. Box 85, Mt. Hamilton, CA 95140;
[d] Lawrence Livermore National Lab., IGPP, L-413, 7000 East Ave., Livermore, CA 94550

## ABSTRACT

At the University of California's Lick Observatory, we have implemented an on-sky testbed for next-generation adaptive optics (AO) technologies. The Visible-Light Laser Guidestar Experiments instrument (ViLLaGEs) includes visible-light AO, a micro-electro-mechanical-systems (MEMS) deformable mirror, and open-loop control of said MEMS on the 1-meter Nickel telescope at Mt. Hamilton. (Open-loop in this sense refers to the MEMS being separated optically from the wavefront sensing path; the MEMS is still included in the control loop.) Future upgrades include predictive control with wind estimation and pyramid wavefront sensing. Our unique optical layout allows the wavefronts along the open- and closed-loop paths to be measured simultaneously, facilitating comparison between the two control methods. In this paper we evaluate the performance of ViLLaGEs in open- and closed-loop control, finding that both control methods give equivalent Strehl ratios of up to $\sim 7\%$ in I-band and similar rejection of temporal power. Therefore, we find that open-loop control of MEMS on-sky is as effective as closed-loop control. Furthermore, after operating the system for three years, we find MEMS technology to function well in the observatory environment. We construct an error budget for the system, accounting for 130 nm of wavefront error out of 190 nm error in the science-camera PSFs. We find that the dominant known term is internal static error, and that the known contributions to the error budget from open-loop control (MEMS model, position repeatability, hysteresis, and WFS linearity) are negligible.

## 1. INTRODUCTION

ViLLaGEs, the adaptive optics (AO) system on the Nickel telescope at Lick Observatory, is an on-sky testbed for vetting next-generation AO components and algorithms. These new technologies include micro-electro-mechanical systems (MEMS) deformable mirrors, open-loop control as envisioned for multi-object adaptive optics (MOAO), predictive control with wind estimation, pyramid wavefront sensing, and up-link laser guide star correction. "Visible-Light Laser Guidestar Experiments" (ViLLaGEs) has been operating at Mount Hamilton since its first light in Fall 2007.[1] In this paper we describe the instrument and analyze its on-sky performance to date, demonstrating MEMS deformable mirrors and open-loop AO control in an authentic observatory environment.

## 2. THE VILLAGES AO SYSTEM

The parameters of the ViLLaGEs AO system are described in Tab. 1. It operates in the visible wavelengths at up to 1000 Hz in either closed- or open-loop control. The wavefront corrector is a 144-actuator continuous facesheet Boston Micromachines MEMS deformable mirror (DM). The wavefront sensor (WFS) is a 10x10 Shack-Hartmann arranged as a 4x4 "hexadecacell" in the Fried geometry (Fig. 1, left). A unique optical layout (Fig. 2) allows for corrected and uncorrected paths to be measured simultaneously. That is, compensated and uncompensated Shack-Hartmann spots are recorded on the WFS camera no matter which loop is controlling AO, and corrected and uncorrected PSFs are imaged on the science camera at all times also. This facilitates comparison between the two control methods, as well as diagnostics of the AO performance.

---

[*]Contact: ktmorz@ucolick.org

Table 1. Parameters of the ViLLaGEs AO system.

| Property | Description |
| --- | --- |
| Location | 1-m Nickel telescope, Lick Observatory, Mt. Hamilton, California |
| Deformable mirror | 144-actuator (84 illuminated) Boston Micromachines MEMS continuous Au facesheet |
| Wavefront sensor | Shack-Hartmann, 60 illuminated subapertures, 11.1-cm subapertures, Fried geometry, 4x4 pixels per subaperture ("hexadecacell") |
| Control | 1 kHz closed-loop or open-loop control |
| WFS camera | SciMeasure Li'l Joe $\lambda \sim 300-1000$nm |
| Science camera | B, V, R, I, 900/40 Facility CCD |

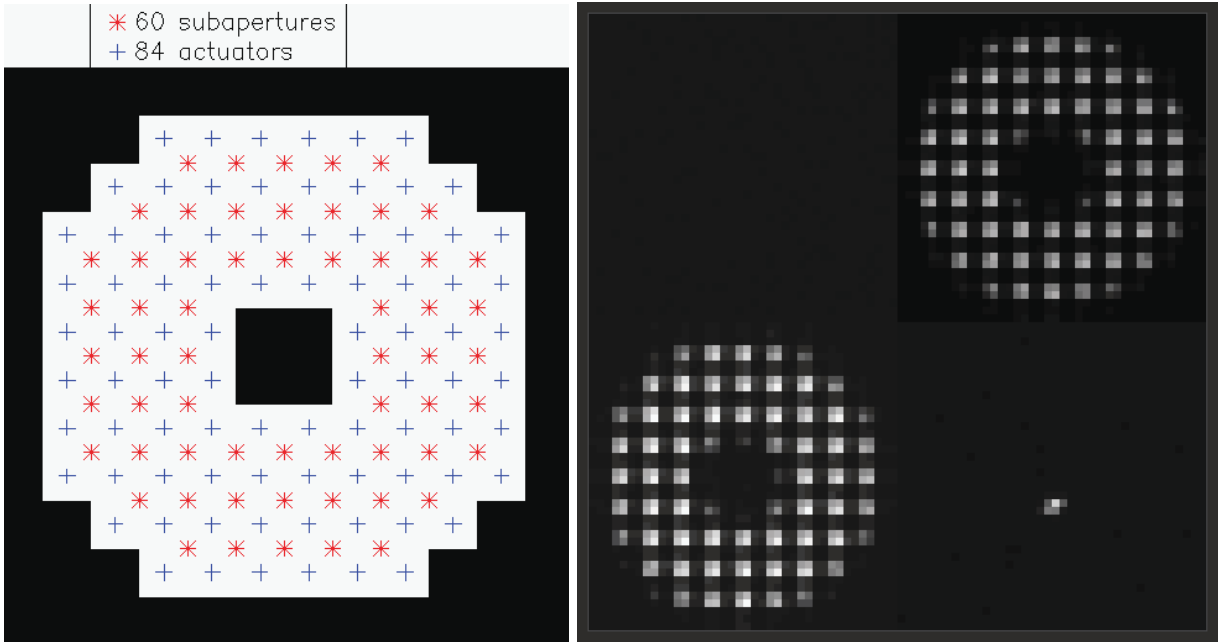

Figure 1. (Left) Illuminated Shack-Hartmann (S-H) subapertures and MEMS actuators. (Right) Raw WFS camera image, internal fiber, flattened. Compensated (top right) and uncompensated (lower left) S-H spots, and tip/tilt pick-off.

## 2.1 Optical layout

Figure 2 shows the AO system layout. The telescope produces an f/17 beam that is relayed to the AO bench mounted at the Cassegrain focus. Downstream of the tip/tilt mirror, half the light is reflected off the MEMS DM for closed-loop control, while the other half bypasses the DM for open-loop control. The wavefront sensor measures both the post-DM (closed-loop AO) and the non-DM (open-loop AO) spots, as well as the tip/tilt image that is re-inserted above the lenslet array (Fig. 1, right). The science camera simultaneously images the corrected and uncorrected PSF.

## 2.2 ViLLaGEs experiments

ViLLaGEs was designed to test several next-generation AO technologies in the observatory environment. These include controlling a MEMS on-sky in closed-loop, controlling AO in open-loop control,[2] wind-predictive control,[3] pyramid wavefront sensing,[4,5] and uplink LGS correction.[6] (It is now likely, however, that the uplink LGS experiment will be tested at the Shane 3-m rather than the Nickel 1-m.) The experiments are outlined in Tab. 2.

## 2.3 The Villages MEMS deformable mirror

The high-order deformable mirror in ViLLaGEs is a 144-actuator (12x12) Boston Micromachines MEMS, shown in Fig. 3. In operation at Lick Observatory, 10x10 of the actuators are illuminated whilst the outer rings are slaved and the 7 actuators at each corner are set to zero volts to conserve communication channels. Before going on-sky with the AO system, we flattened the MEMS in the LAO in September 2007 to 12 nm rms (6 nm in-band) over a 10-actuator-diameter pupil using a Zygo interferometer. We saved the voltages that flatten the MEMS, and apply them when at the telescope for alignment or reference calibration purposes along the MEMS path (e.g., rather than using the unpowered shape, with its $1.2\mu$m peak-valley curvature). The influence function of this particular MEMS was measured and found to be 2 actuators full-width at half-max (Fig. 4). We also calibrated the MEMS model with the Zygo in the LAO before transporting it to the telescope (see § 2.4).

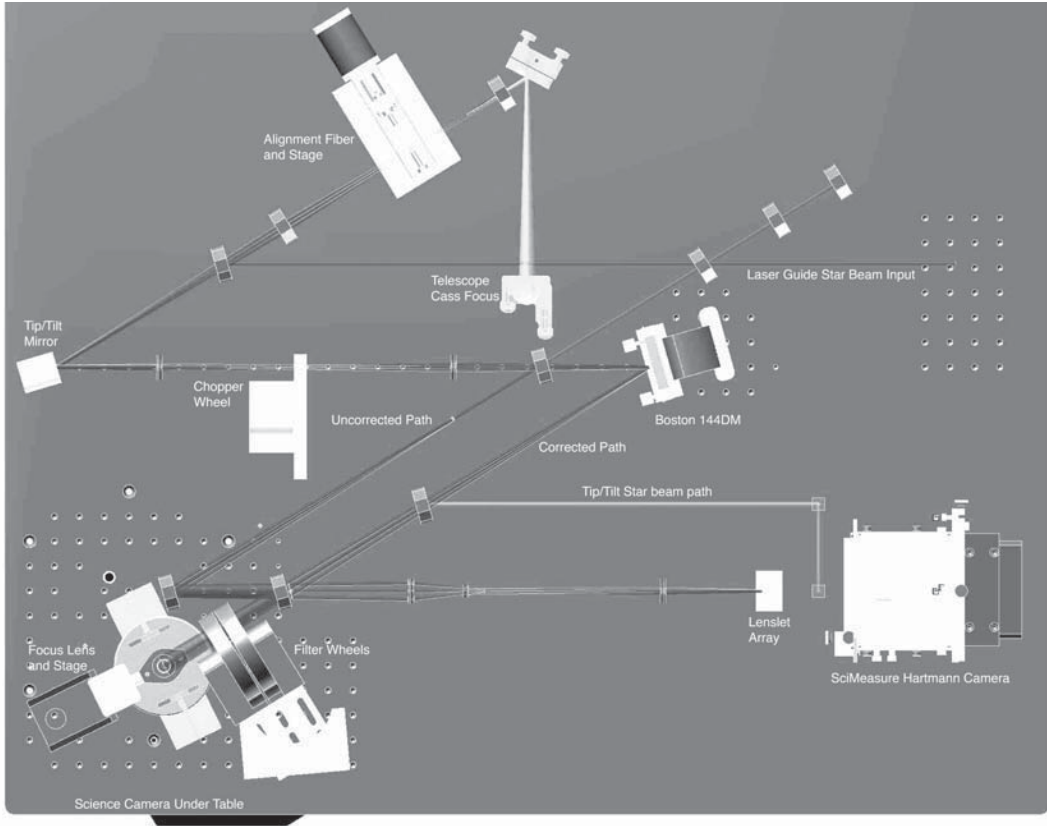

Figure 2. Layout of the ViLLaGEs AO system. The wavefront sensor is at lower right and the science camera is at lower left (into the page). The PI tip/tilt mirror is common to both optical paths, whereas the MEMS high-order deformable mirror is not. Optical paths including the MEMS are the closed-loop Shack-Hartmann WFS and the corrected PSF on the science camera; paths bypassing the MEMS are the open-loop WFS and the uncorrected PSF on the science camera.

Table 2. ViLLaGEs on-sky experiments.

| Phase I | Completed |
|---|---|
| 1. Control MEMS-AO in closed-loop | Fall 2007 |
| 2. Control MEMS-AO in open-loop | Spring 2008 |
| **Phase II** | **Target** |
| 3. Pyramid wavefront sensing | Fall 2010 |
| 4. Uplink LGS correction | Lick 3m, 2011-12 |

### 2.4 MEMS model and open-loop control

MEMS DMs are position-repeatable[7] and lack hysteresis[8] at the sub-nanometer level; they are therefore ideal for open-loop control in which the WFS path bypasses the DM. Open-loop control on ViLLaGEs is achieved via a phase-to-volts look-up table for the MEMS commands. The wavefront sensor records the open-loop centroids, the reconstructor calculates the phase of the wavefront, and the MEMS commands are determined from a look-up table, as shown in Fig. 5. The phase-to-volts table is derived from a semi-empirical model of the MEMS.

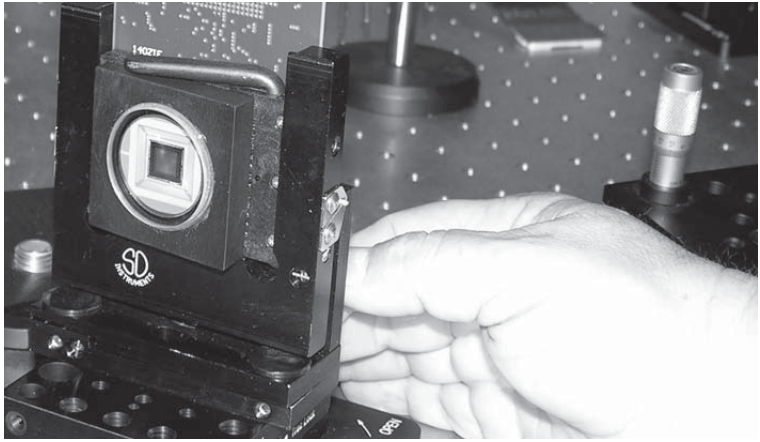

Figure 3. The ViLLaGEs 144-MEMS from Boston Micromachines.

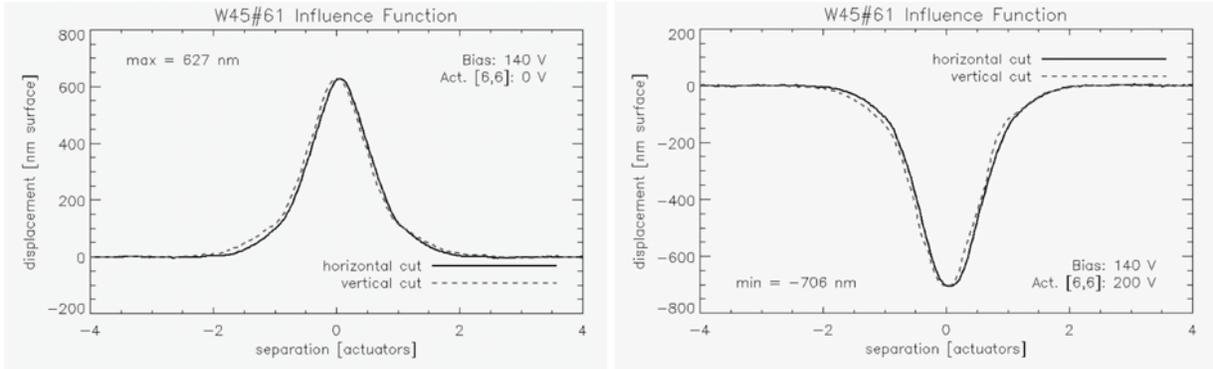

Figure 4. Influence function of the ViLLaGEs MEMS. FWHM = 2 actuators.

This model works as follows: The MEMS is modeled as a thin plate, such that its surface is described by the thin-plate equation:

$$\nabla^4 w_z = f_p/D \tag{1}$$

where $w_z$ is the displacement of the surface in the normal direction, $f_p$ is the net plate force on the top sheet, and $D$ is the flexural rigidity, a constant that depends on the material of the plate. The net force is zero for a non-accelerating plate, and is composed of the spring force and the electrostatic force:

$$f_p(V,d) = f_s(d) + f_e(V,d) = 0. \tag{2}$$

In this equation $f_s \propto -d$ is the restoring spring force of the actuator and plate, a function of displacement $d$; $f_e \propto V/d^2$ is the electrostatic force, a function of displacement and voltage $V$. The model is calibrated empirically by measuring the force on an actuator at each combination of voltage (in 20-volt increments) on that actuator and its neighbors. Details of the model and results in the lab are given in a previous LAO work.[2]

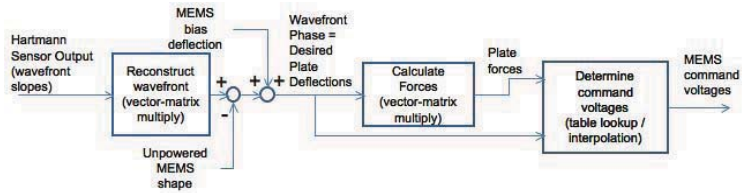

Figure 5. Block diagram of the open-loop control process.

## 3. VILLAGES PERFORMANCE

In this section we analyze the performance of the ViLLaGEs AO system at the telescope, in closed- and open-loop control. Telemetry data are used to calculate the error budget. The resultant wavefront error is compared to a ViLLaGEs AO model and to stellar PSFs.

Telemetry diagnostics consist of streams of data recorded at the frame rate of the AO system, which in this paper is always 1000 Hz for both closed- and open-loop control. The data that are available consist of: the raw WFS camera images (1000 samples), the intensities per subaperture (4096 samples), the centroids (4096 samples), and the tip/tilt residuals (1024 samples). Science camera images were recorded simultaneously with the telemetry data.

### 3.1 Science camera images

Figure 6 illustrates the similar performance of closed- and open-loop control. $\gamma$ Lyr (V$\sim$ 3) is imaged in sequential 30-second exposures, first controlled closed-loop and then controlled open-loop. Both images are typical with 7% Strehl. Figure 7 shows $\alpha$ Her (V$\sim$ 3) in I and R-band controlled in open-loop AO. In all science camera images the upper-right PSF is from the uncorrected path while the lower-left PSF is the post-MEMS corrected image; they are separated 12" from each other. In most images the uncorrected image has been scaled to equalize the throughput along the two paths. North is down and East is right. The plate scale is 0.029"/pix.

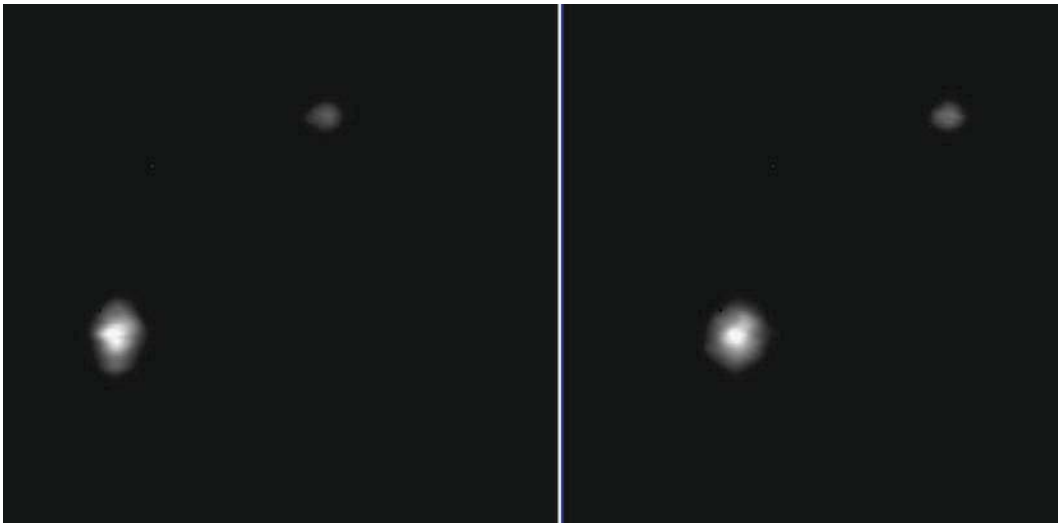

Figure 6. Thirty-second exposures of $\gamma$ Lyr, log scale (June 2010). Left: closed-loop AO; 7% I-band Strehl. Right: open-loop AO; 7% I-band Strehl. Within each image, upper-right is the uncorrected and lower-left is the corrected PSF.

The best Strehl ratio consistently achieved with ViLLaGEs in recent months is $\sim$ 7% in I-band, the same for both closed- and open-loop control. Table 3 summarizes the image quality achieved in May and June 2010. 7% Strehl in I-band corresponds to 210 nm wavefront error (WFE) according to the Marechal approximation.[9] It is the object of this section to use the AO telemetry data to determine the WFE distribution, or error budget.

### 3.2 Turbulence parameters

Table 4 lists the turbulence parameters calculated (below) from telemetry data, following Hardy chapter 9.[9]

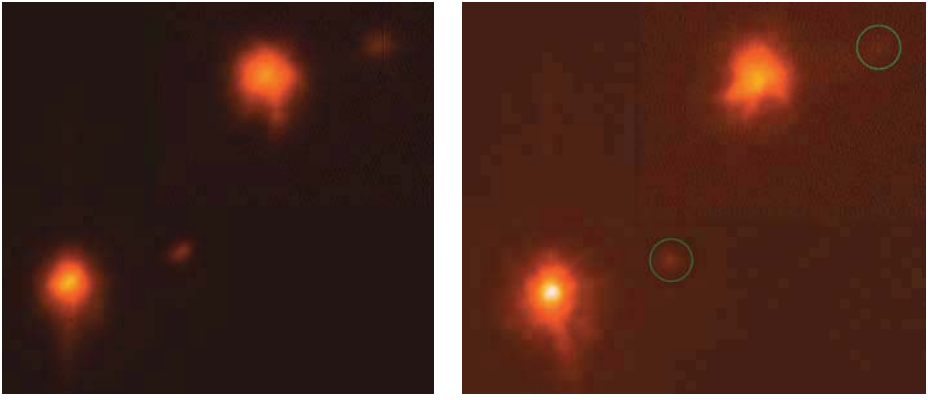

Figure 7. Binary $\alpha$ Her (itself a 0.3″ close binary of V~3.1 and 3.5) and HR 6407 (V~5.4), separated by ~ 5″. Open-loop AO, log-scale images, May 2010. Left: R-band, 2.5-second exposure. Right: I-band, 0.75-second exposure.

Table 3. Image quality, May 2010, $\alpha$ Her (Fig. 7).

| Filter[10] | Integration Time / s | Uncorrected FWHM | AO-corrected FWHM | Strehl |
|---|---|---|---|---|
| B | 180 | 1.24″ | 0.76″ | 1% |
| V | 5 | 1.27″ | 0.70″ | 2% |
| R | 2.5 | 0.78″ | 0.57″ | 3% |
| I | 0.75 | 0.78″ | 0.51″ | 7% |

Table 4. Turbulence parameters.

| | |
|---|---|
| $r_0$ | 12 cm |
| $v$ | 10 m/s |
| $f_G$ | 40 Hz |
| $\tau_0$ | 4 ms |

### 3.2.1 Spatial coherence

The Fried parameter, $r_0$, is the length scale over which atmospheric turbulence is coherent to 1 rad$^2$ rms. There are various ways to measure $r_0$. Here we use the centroids from the uncompensated wavefront sensor to reconstruct the full high-order atmosphere. $r_0$ is then determined from the reconstructed phase according to

$$r_0 = D\left(\frac{\alpha_c}{\sigma_\phi^2}\right)^{3/5} \qquad (3)$$

where $D$ is the diameter of the telescope primary, $\alpha_c = 0.14$ is the fitting coefficient for a circular segment with piston and tip/tilt removed, and $\sigma_\phi$ is the rms WFE.[9] The median $r_0$ at $\lambda = 500$ nm was **12 cm** and ranged from 8 to 15 cm. These values were corroborated by measuring the FWHM$\approx \lambda/r_0$ of the uncompensated PSFs on the science camera.

### 3.2.2 Temporal coherence

We estimate the wind speed from the tip/tilt temporal power spectra (Fig. 8). The mid-frequency power goes as $f^{-8/3}$, while the power at low frequencies goes as $f^{-2/3}$ for the tilt-included and as $f^{4/3}$ for the tilt-removed phase.[9] From the plot, we estimate the slope changeover to occur around 7 Hz. The wind velocity can then be estimated according to $v = f_c D/0.705 \approx$ **10 m/s**.[9] The Greenwood frequency is then given by $f_G = 0.427v/r_0 =$ **40 Hz** and the temporal coherence $\tau_0 = 0.134/f_G \approx$ **4 ms**.[9]

### 3.3 Error budget $\sigma_\phi$

We now construct an error budget for ViLLaGEs under open-loop control (Tab. 6). Similar efforts for closed-loop AO systems have been undertaken by Troy,[11] van Dam,[12,13] and many others, but open-loop error budgets are somewhat more straightforward in that the full reconstructed wavefront can be utilized for calculations, as it is available from the uncompensated WFS data. Furthermore, the high 1-kHz frame rate of ViLLaGEs being within the temporal coherence provides for alternative calculation of measurement and temporal error, as we will see below. The error terms investigated are: tip/tilt, fitting, temporal, measurement, internal static, calibration, linearity, and the three terms unique to open-loop control. The total wavefront error is then compared to the Strehl in a contemporaneous stellar PSF using the extended Marechal approximation.[9]

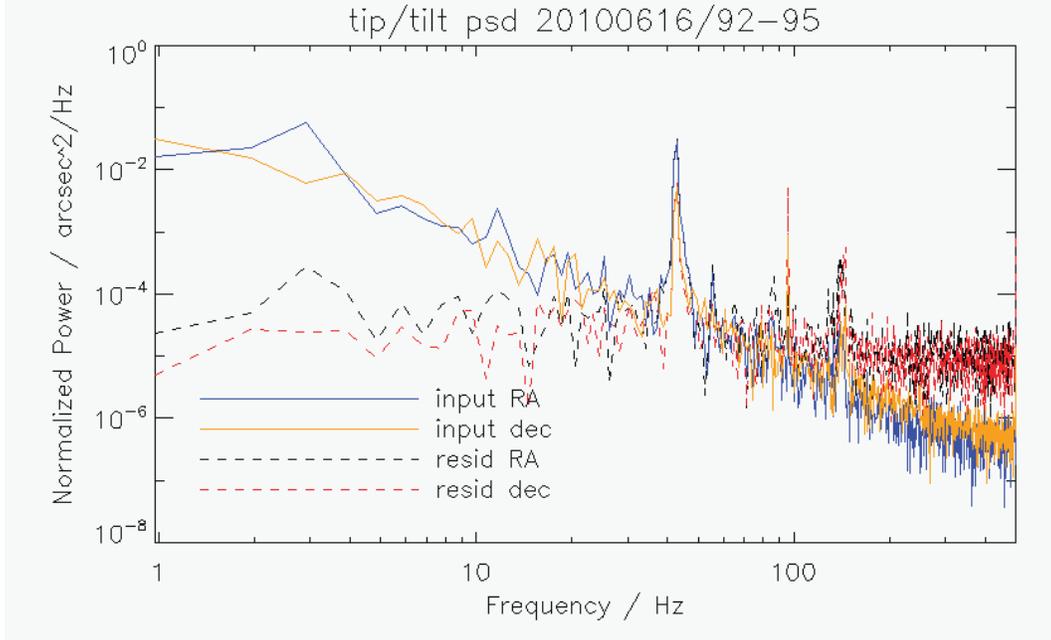

Figure 8. Temporal power spectrum, tip/tilt ($\gamma$ Lyr, Fig. 6). Tip/tilt error is approximately 300 mas or 80 nm rms. The 40-Hz vibration has only been present recently and its origin is as yet unknown.

### 3.3.1 Tip/tilt error $\sigma_{t/t}$

The tip/tilt error (Fig. 8) is estimated from the rms of the difference between wavefronts reconstructed with and without tip/tilt. It ranged from **73-88 nm rms**, or **270-330 mas**. To find the remaining high-order terms in the error budget, the tip/tilt error was subtracted in quadrature from the total wavefront error.

### 3.3.2 Fitting error $\sigma_f$

We calculate the fitting error using the model of a DM as a spatial filter that passes spatial frequencies higher than its actuator spacing. Therefore we use the formula

$$\sigma_f^2 = \alpha_F \left(\frac{d}{r_0}\right)^{5/3} \qquad (4)$$

where $d = 11.1$ cm is the subaperture size and $\alpha_F$ is the fitting coefficient[9] — we use $\alpha_F = 0.38$ to account for aliasing. We find the fitting error to be **45±10 nm rms** for the range of $r_0$'s.

### 3.3.3 Temporal error $\sigma_t$

Temporal error for open-loop control can be estimated from the temporal power spectra assuming a transfer function of unity for open-loop AO. Figure 9 shows the open-loop power spectrum. (The closed-loop temporal power spectrum is shown for comparison in Fig. 10. Note that the rejection is similar for open- and closed-loop AO, implying both control methods work equally well.) The rectangular area obtained by extending the noise floor from the high frequencies across to the low delineates the measurement noise and is in agreement with the value from §3.3.4. The remaining area above the noise floor and below the open-loop-AO-corrected power is the temporal noise. This method of integrating the temporal power spectrum of the reconstructed wavefronts gives **55 nm rms** temporal error in open-loop control.

This approximately agrees with the value found using the time lag $\tau_S = 2$ ms for our two-step delay system and the temporal coherence $\tau_0 \approx 4$ ms with the formula[9]

$$\sigma_t^2 = \left(\frac{\tau_S}{\tau_0}\right)^{5/3} \qquad (5)$$

giving $\sigma_t = 45 \pm 10$ nm, if we assume $\pm 1$-ms error on $\tau_0$.

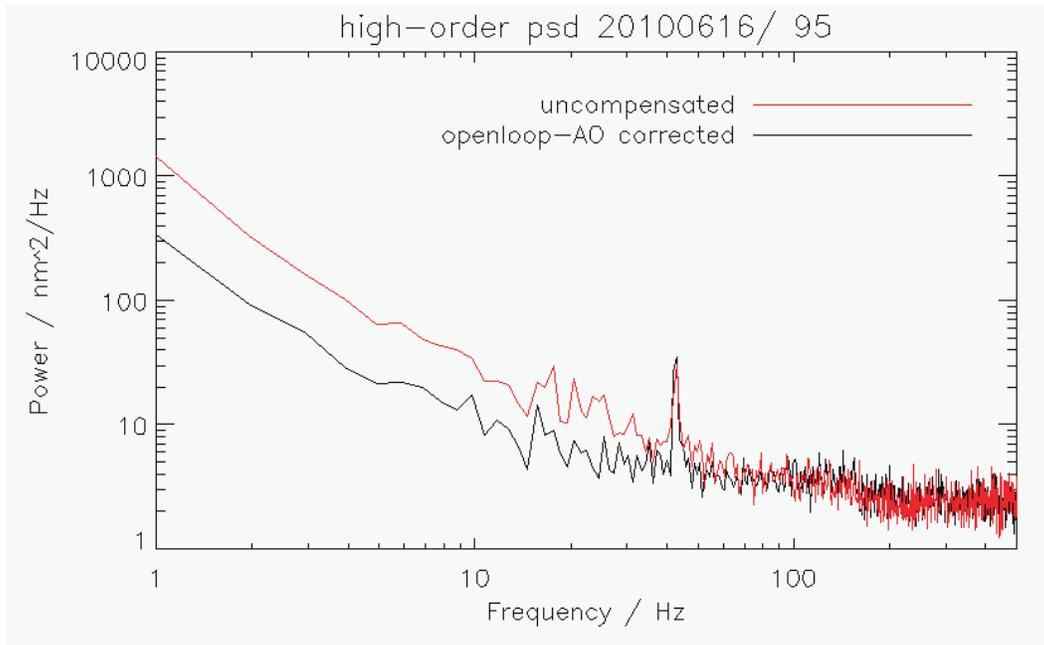

Figure 9. High-order temporal power spectrum, open-loop control ($\gamma$ Lyr, Fig. 6).

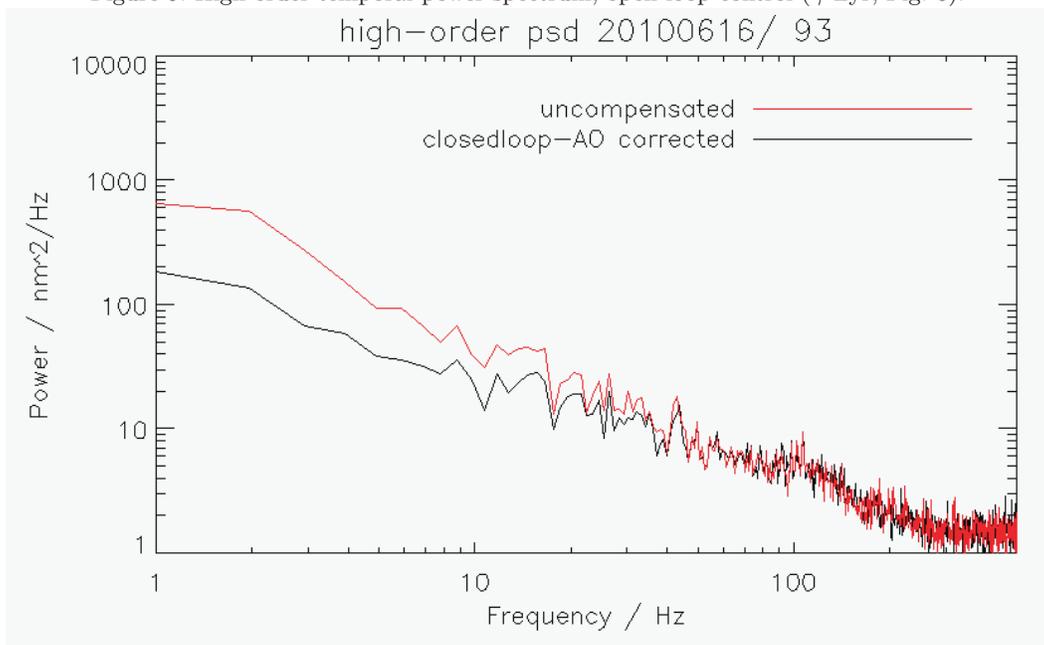

Figure 10. High-order temporal power spectrum, closed-loop control ($\gamma$ Lyr, Fig. 6). Both PSDs are made with tip/tilt-removed reconstructed wavefronts, but some residual 40-Hz vibration (Fig. 8) is seen in the open-loop-control PSD.

### 3.3.4 Measurement error $\sigma_m$

Measurement error, $\sigma_m$, is the error due to inaccuracies in measuring the wavefront, including photon noise, read noise and dark current on the WFS camera, as well as errors in the reconstruction. We use the centroids from the uncompensated WFS to reconstruct the wavefront independently off-line using fast Fourier wavefront reconstruction,[14] and take the difference images to get the measurement noise.

The difference between successive wavefronts, $\Delta\phi_m$, is given by $\Delta\phi_m = \phi_m^t - \phi_m^{t+1}$, where $\phi_m^t$ is the measured wavefront at iteration $t$. The measured and reconstructed wavefront, $\phi_m$, includes the noise in the measured wavefront, $n_\phi$. At 1000 Hz the frames are spaced 1 ms apart. If we therefore assume that the wavefront due to turbulence in the atmosphere, $\phi_a$, is constant over the course of 1 ms — which is well within the temporal coherence of $\tau_0 \approx 4$ ms — then the variance of the difference images $\Delta\phi_m$ gives the measurement noise as follows:

$$
\begin{aligned}
\phi_m^t &= \phi_a^t + n_\phi^t \\
-\left(\phi_m^{t+1}\right. &= \left.\phi_a^{t+1} + n_\phi^{t+1}\right) \\
\phi_m^t - \phi_m^{t+1} &= \phi_a^t - \phi_a^{t+1} + n_\phi^t - n_\phi^{t+1} \\
\Delta\phi_m &= n_\phi^t - n_\phi^{t+1} \\
&= \sqrt{2} n_\phi \\
\therefore \sigma_m &= rms(\Delta\phi_m/\sqrt{2})
\end{aligned}
\tag{6}
$$

The average rms of ten thousand difference images from 10 streams of telemetry data is $\Delta\phi_m/\sqrt{2}$ is $\sigma_m =$ **40 ± 5 nm rms**. The value was verified by following the variance of one actuator's phase difference in time.

A double-check for measurement error can be carried out by integrating the temporal PSD as done in §3.3.3. The area in the PSD (Fig. 9) under an imaginary horizontal line extended from the noise floor at high frequencies across to the low frequencies gives a value of 37 nm for the measurement noise, in agreement with the value from the wavefront differences.

We can calculate the measurement error a third way: The intensity measured on the uncompensated WFS for a 3rd-magnitude star is ~1800 counts/subaperture, and with a negligible dark current and a read noise of 10 e$^-$/pix and 16 pixels/subap. the signal-to-noise ratio SNR $\approx$ 31. (Comparing the measured counts to the known magnitude of the star gives the WFS CCD throughput as 37% for both spot patterns.) Putting this SNR into the formula given in Hardy chapter 5[9]

$$\sigma_m = \sqrt{2}\frac{\pi^2 K_g}{4(SNR)}\left[\left(\frac{3}{2}\right)^2 + \left(\frac{\theta d}{\lambda}\right)^2\right] \tag{7}$$

gives $\sigma_m = 27$ nm, where $K_g = 1.3$ is a constant to account for centroiding errors due to fill factor on the CCD, $\theta = 1.6$" is the spot size (§4), $\lambda = 500$ nm is the wavelength, and with a noise-propagator of unity for open-loop control. Therefore we increase the error bars on measurement error and note that the value of 40 nm is measurement + reconstructor error, plus any residual temporal error below 1 kHz frequency.

Finally, we investigate temporal and measurement error in combination, as the methods we use to calculate them are intertwined. The measurement + temporal error using the area under the PSD are 37 nm + 53 nm = 65 nm in quadrature. The measurement + temporal error using equations 5 & 7 are 27 + 45 = 52 nm. Therefore these values are within the ~ 10-nm error bars. As the theoretically-based equations 5 & 7 do not account for reconstructor or other errors, we take instead the values from the empirically-based PSD-integration method for Tab. 6.

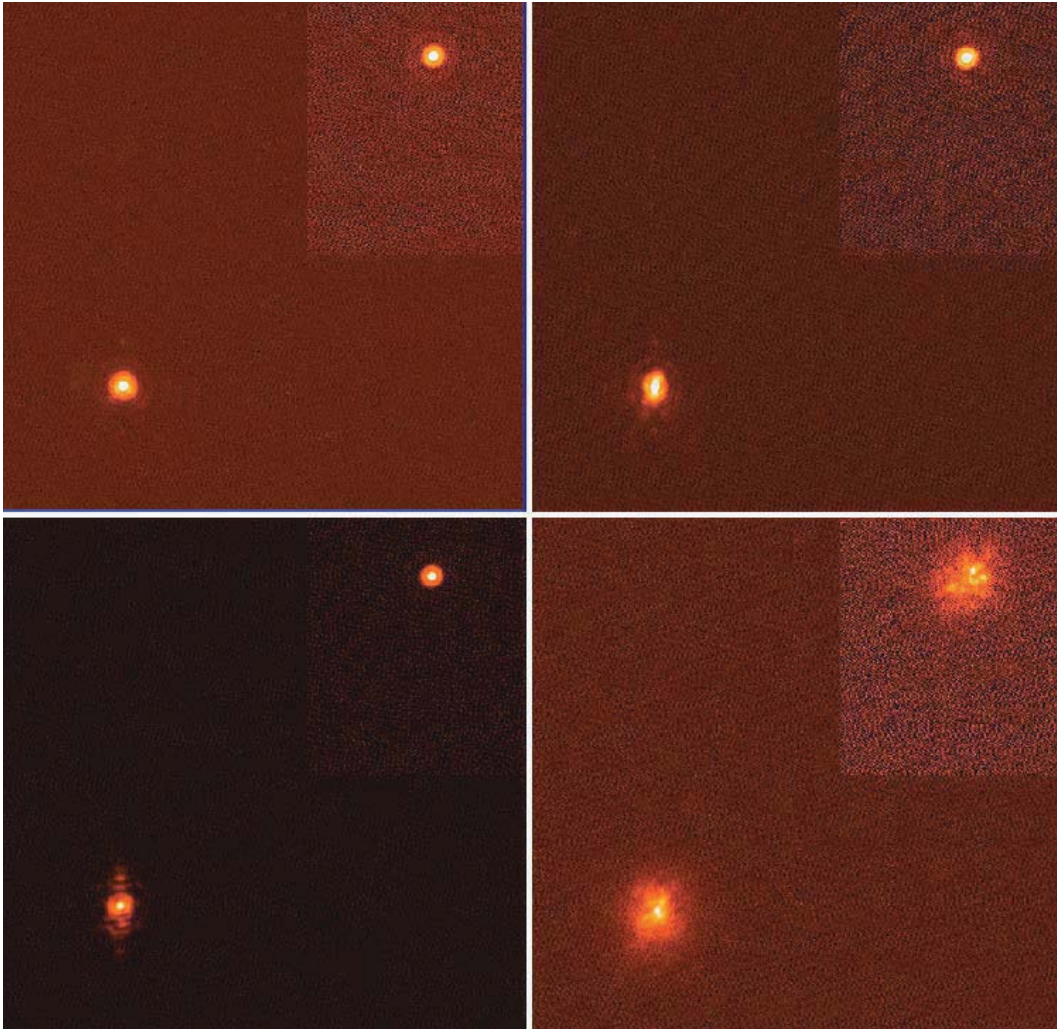

Figure 11. Internal fiber, log scale. (Upper left) No AO correction. (Upper right) Open-loop AO correction. (Lower left) Closed-loop AO correction. (Lower right) Open-loop AO correction with rotating Kolmogorov phase screen.[15]

### 3.3.5 Internal static $\sigma_{int}$ and calibration error $\sigma_{cal}$

A tungsten white-light source placed in the focal plane upstream of the tip/tilt mirror is used to measure the internal error. When using the fiber, an obscuration modeling the Nickel secondary is inserted upstream of the tip/tilt mirror. The internal static error is obtained by flattening the MEMS to the Zygo-determined flat voltages and measuring the wavefront error in the PSF to be $\sigma_{int}$ = **90 nm rms** (Fig. 11, top left).

Calibration error is calculated by measuring the Strehl with the DM AO-controlled (Fig. 11, top right and bottom left). This error term is due to calibration errors, as new reference centroids were saved immediately before conducting this test. Table 5 lists the image quality obtained with the fiber, as compared to the Airy PSF, for the following cases: non-DM path, DM path without AO, and DM path with AO correction. There is a Strehl degradation in going from Zygo-flattened MEMS to AO-controlled MEMS (open-loop control and closed-loop control give approximately the same Strehl degradation, though the images differ in structure). We have seen this before on the ExAO (extreme AO) testbed at the LAO and with Lick AO, and the effect is small but non-trivial to remove; it is noted here for completeness. The wavefront error difference from the Zygo-flat PSF to the AO-controlled PSF on a bare fiber is $\sigma_{cal}$ = **40 nm rms**.

Table 5. Image quality of internal fiber ($\lambda = 635$ nm) (Fig. 11).

| PSF | Uncorrected FWHM | Uncorrected Strehl | AO-corrected FWHM | AO-corrected Strehl |
|---|---|---|---|---|
| Non-DM path | 0.142" | 70% | | |
| DM path | 0.149" | 45% | 0.153" | 40% |
| Turbulence generator | 0.626" | 10% | 0.307" | 15% |
| Airy | | | 0.134" | 100% |

### 3.3.6 WFS linearity $\sigma_{lin}$

Ammons et al.[16] calculate the error due to WFS deviations from linearity (a function of Shack-Hartmann spot size) to be **15 nm rms**, using analytically-determined spot sizes. We verify this calculation below (§4) with an empirical determination of the spot size to be 1.6". Corroboration of this small value for WFS linearity is seen in the nearly-linear curves of Fig. 13.

### 3.3.7 Open-loop AO errors $\sigma_{model}$, $\sigma_{goto}$, $\sigma_{hyst}$

The error terms unique to open-loop control are the error in the MEMS phase-to-volts model $\sigma_{model}$, the error in the MEMS position repeatability (a.k.a. go-to capability of the actuators) $\sigma_{goto}$, and any other positioning error such as hysteresis $\sigma_{hyst}$.

The model error is a measure of how accurately the desired phase is applied to the MEMS, given that the voltage commands are calculated using the model described in §2.4. It was measured with the Zygo interferometer in the LAO by applying Kolmogorov phase screens to the MEMS using the model, and measuring the rms deviation to be 15 nm for an $r_0$ of approximately 10 cm. This error increases with increasing wavefront amplitude.[2]

The go-to capability is a measure of how well the actuators go to a set place in nm phase when given a known command in volts. It was measured on the ExAO testbed with a phase-shifting diffraction interferometer (PSDI) by commanding the actuators repeatedly away from and then back to the same voltage, and was found to be 0.05 nm.[7] The hysteresis was measured by driving the actuators through a loop of progressively increasing and then decreasing voltage, and was found to be 0.3 nm.[8]

### 3.4 Total wavefront error

Table 6. High-order error budget for open-loop AO.

| Error source | $\sigma_{WFE}$ | Data nm | ± | Model nm |
|---|---|---|---|---|
| DM fitting | $\sigma_f$ | 45 | 10 | 44 |
| Temporal | $\sigma_t$ | 55 | 10 | 44 |
| Measurement | $\sigma_m$ | 40 | 15 | 28 |
| Internal static | $\sigma_{int}$ | 90 | 5 | ⊘ |
| Calibration | $\sigma_{cal}$ | 40 | 5 | ⊘ |
| WFS linearity[16] | $\sigma_{lin}$ | 15 | 5 | ⊘ |
| MEMS model[2] | $\sigma_{model}$ | 30 | 10 | ⊘ |
| MEMS position-repeatability[7] | $\sigma_{goto}$ | 0.05 | 0.02 | ⊘ |
| MEMS hysteresis[8] | $\sigma_{hyst}$ | 0.3 | 0.1 | ⊘ |
| **Total Calculated WFE** | $\sigma_{tot}$ | **130** | **25** | **68** |
| Measured WFE (from PSF) | $\sigma_{PSF}$ | 190 | 10 | |
| *Unaccounted WFE* | $\sigma_?$ | *140* | *25* | |

The total error budget (Tab. 6) is found by summing the individual error sources in quadrature.

$$\sigma_{tot} = \sqrt{\sigma_f^2 + \sigma_t^2 + \sigma_m^2 + \sigma_{int}^2 + \sigma_{cal}^2 + \sigma_{lin}^2 + \sigma_{model}^2 + \sigma_{goto}^2 + \sigma_{hyst}^2}$$
$$= 130 \pm 20 \text{ nm} \quad (8)$$

We account for 130 nm of high-order error. The 7%-Strehl I-band PSF indicates 190 nm of high-order WFE. The missing error amounts to 140 nm.

### 3.4.1 Modeled WFE

The fitting, temporal, and measurement error are compared to an end-to-end IDL model of the ViLLaGEs system. The model was computed with the following parameters: NGS $V = 3$, wind speed $v = 10$ m/s, $r_0 = 12$ cm, $K_g = 1.3$, 30% WFS CCD throughput, 80% WFS CCD quantum efficiency, spot size = 1.6", and read noise 10 $e^-$/pix. Fitting error in the model is calculated empirically by differencing the input wavefront and the DM shape, generated using superposition of 2-actuator-wide gaussian influence functions. Measurement error is calculated with Eqtn. 7. Temporal error is the residual error left after subtracting off fitting error in a case with no measurement noise.

Fitting error shows excellent agreement with the data (Tab. 6). Temporal and measurement error are underestimated by about 10 nm in the model as compared to the empirically-based/PSD-derived values of 55 and 40 nm, respectively, for the data in Tab. 6. However, these modeled values do agree with the theoretically-based/telemetry-derived values of 27 and 45 nm, respectively, for temporal and measurement error from Eqtns. 5 & 7.

The measurement error term in the data also includes reconstructor error, which may, along with uncertainty, account for the discrepancy. Nevertheless, the ±10 nm consistency between the data and the model in these three terms implies that the missing WFE is coming from some unmodeled term such as internal calibration or open- or closed-loop control errors such as reconstruction.

### 3.4.2 Unaccounted WFE

As we speculate as to the source of the missing 140 nm WFE, we note a similar magnitude of unaccounted error in the literature.[12] Our dominant known source is internal static error, so we plan to implement an image-sharpening routine. Thus alignment and registration of the MEMS to the compensated WFS, and the uncompensated WFS to the MEMS, are important. There may be clipping or vignetting at the field stop, affecting the linearity of the WFS measurements at high dynamic ranges. Flexure might be a consideration worth exploring in the future — we normally stay within a few hours of zenith, but we have had to re-secure the DM and WFS mounts for stability. Another subtlety is the influence of the 7 actuators at each outer corner being set to 0 volts to conserve bandwidth. However, the influence function is narrow enough and the pupil inscribed within a small enough region from the edges that this is negligible, as is confirmed by varying the bias when closing to the internal fiber and seeing no change in Strehl.

As the WFS camera is broad-band, the spot size may vary from 0.6–1.9" (§4) (i.e. the diffraction-limit across the full quantum-efficient spectrum of the CCD) or more with seeing. A smaller spot size than the 1.6" used here will tend to reduce the measurement error, but will increase linearity error. However, these effects seem negligible in that these error terms are among the smaller sources. The spot may be chromatic, but this would be a second-order effect. However, a related source to investigate is the variation in the reference centroid offsets — an average gain is used for the WFS, but the variance in the gain per subaperture, if significant, would be contributing to reconstruction error.

In an effort to understand the system and determine if an uncertain and variable spot size could be contributing to the unknown wavefront error, we conducted an experiment to measure the spot size (§4).

## 4. SHACK-HARTMANN SPOT SIZE

To measure the size of the S-H spot on the wavefront sensor camera, we translated a tungsten while-light source in the focal plane upstream of the tip/tilt mirror (Fig.2). The tip/tilt mirror was held steady and the MEMS was held at the Zygo-flat voltages. The previously-determined plate scale on the science camera (0.029"/pix) gave the distance the fiber moved. The intensities in the WFS were measured and the fiber was stepped until one of the compensated or uncompensated S-H spot patterns lost all flux due to clipping of the fieldstop.

Figure 12 shows the intensity in the two central pixels of each hexadecacell as a function of the position of the fiber. Following the method of Schoeck et al., we model the spot as a gaussian with standard deviation $\sigma$:

$s(x) = \frac{1}{\sqrt{2\pi}\sigma} exp\left(\frac{-x^2}{2\sigma^2}\right)$ and the pixel as a top hat of width $w$.[17] Then the intensity in the pixel is given by the convolution (denoted $\star$) of the spot with the pixel:

$$i(x) = s(x) \star p(x) = \frac{1}{2w}\left[\text{erf}\left\{\frac{1}{\sqrt{2\pi}\sigma}\left((x-x_0)+\frac{w}{2}\right)\right\} - \text{erf}\left\{\frac{1}{\sqrt{2\pi}\sigma}\left((x-x_0)-\frac{w}{2}\right)\right\}\right] \quad (9)$$

By fitting this function to Fig. 12, we find a spot size of 1.6". The plate scale was determined by this fit and independently by the peak-to-peak separation in maximum intensity as the fiber was translated. The mean plate scale was thus found to be 2.13"/pix for the compensated and 1.73"/pix for the uncompensated WFS.

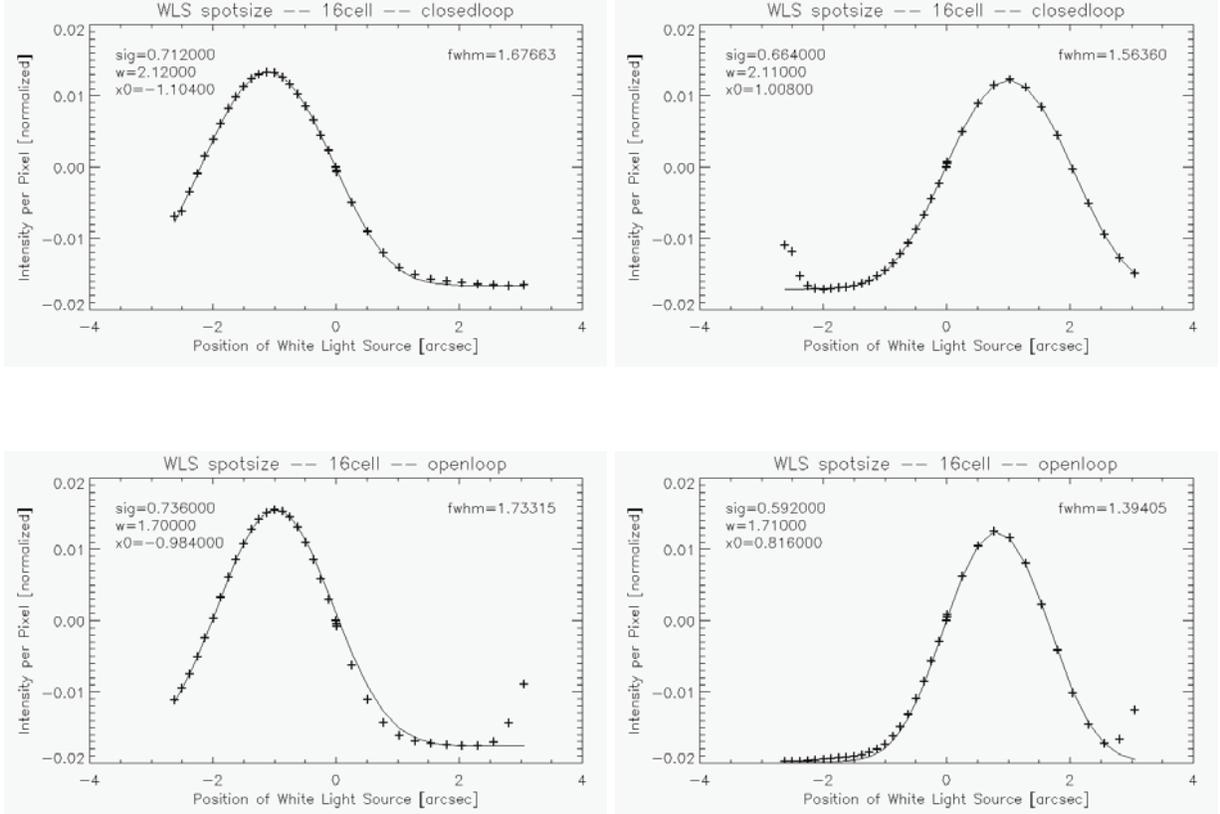

Figure 12. Results of spot size experiment: Intensities in the central two pixels of the hexadecacell, as a function of fiber offset. (Left and right curves correspond to left and right pixels, respectively). Data are fit to Eqtn. 9. Compensated WFS (top): platescale $w = 2.13$"/pix. Uncompensated WFS (bottom): platescale $w = 1.73$"/pix.

We next compute the expected centroid output given the spot size and WFS plate scales, using the weights of the hexadecacell, as follows:

$$C = \frac{-1.5I_0 - 0.5I_1 + 0.5I_2 + 1.5I_3}{I_0 + I_1 + I_2 + I_3} \quad (10)$$

These theoretically-computed centroids are plotted against the center-of-mass data in Fig. 13. The red asterisks are the raw output of the centroider, and a scale factor of 2.3 converts from centroider units to pixels. Once this conversion factor was known, it allowed for accurate reconstruction of the wavefronts from raw centroid data, and has been used for all calculations in §3.3 based on reconstructed wavefronts. The orange plusses in Fig. 13 show the uncompensated WFS centroid output is close to linear across the central pixels of the hexadecacell.

The measured spot is diffraction-limited for our 11.1-cm subapertures at $\lambda = 860$ nm. The spectrum of the tungsten light source rises toward $1\mu$m and the quantum efficiency of the WFS CCD is over 10-20% from 300–1000nm. A smaller spot (e.g., for a star peaked in V-band, or for momentary good seeing) would have a shallower slope. The reference centroid offsets and centroider output are therefore sensitive to the spot size. Therefore the next step is to measure the spot size for a different peak-wavelength source. Furthermore, the spot size will not only vary with peak wavelength but will also vary with seeing.

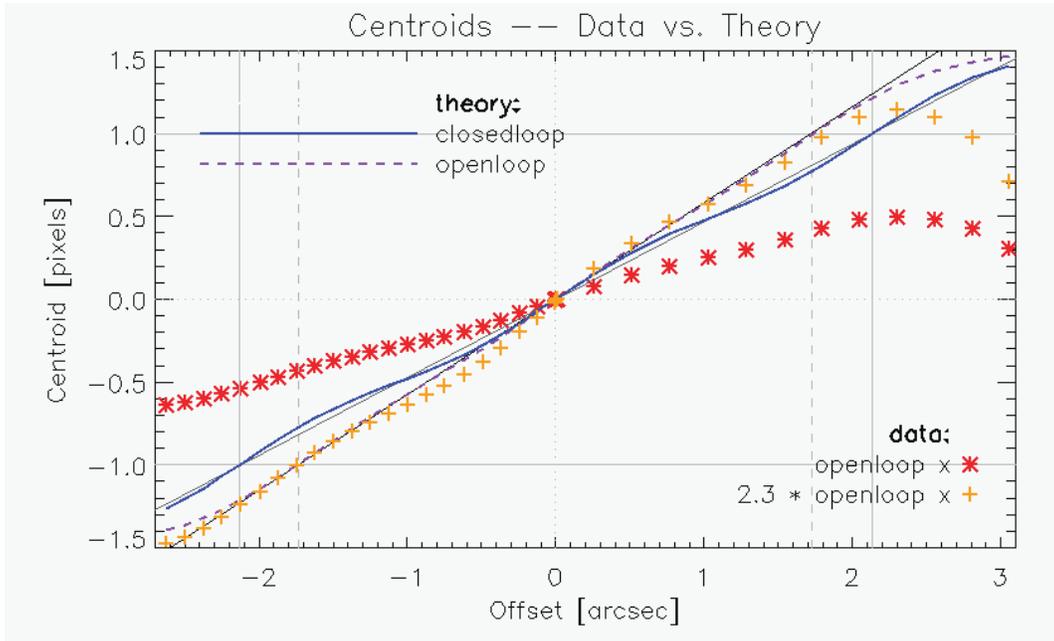

Figure 13. Expected centroids for the measured spot size and plate scale. Theoretical curves (purple dashed: open-loop WFS and blue solid: closed-loop WFS) are approximately linear across the inner two pixels of the hexadecacell, as is data from uncompensated WFS (orange plusses). Vertical grey lines denote pixel edges (solid: closed-loop WFS; dashed: open-loop WFS).

## 5. CONCLUSIONS

In this paper we have presented and analyzed the performance of ViLLaGEs, a MEMS-based visible-light AO system at Lick Observatory operating in closed- and open-loop control. We have been operating MEMS AO on-sky since 2007, and have thus proven MEMS deformable mirror technology in the observatory environment. The laboratory-derived voltages that flattened the MEMS in 2007 still flatten it to 90 nm WFE, including approximately 70 nm internal static error as measured on the non-MEMS path (Fig. 11, top left). This MEMS has a window and care is taken to leave it unpowered when not operating, and thus no actuators have been observed to be damaged.

The closed- and open-loop data show the same performance. Their Strehls are equivalent (up to 7% Strehl in I-band, Fig. 6) and the rejection in the temporal power spectra are similar as well (Figs. 9-10). Therefore, open-loop control is a proven technology on-sky and may be well-considered for certain next-generation AO architectures such as MOAO.

The error budget for open-loop AO tallied to 130 nm rms, whereas the high-order WFE measured at the science camera PSF was 190 nm rms; hence, 140 nm rms went unaccounted. It will likely prove fruitful to discover the source of the additional wavefront error.

One avenue explored for additional error was the spot size, as it could range from a diffraction-limited 0.6" to 1.9" for the broad-band WFS CCD. The spot size was found to be 1.6" (the diffraction limit for $\lambda = 860$ nm) for an internal white-light source that peaks toward the near-IR. The slope of the centroid output as a function of the fiber position in the focal plane is dependent upon spot size, and must be explored further as we attempt to better understand our AO system.

Thus MEMS and open-loop control are proven technologies on-sky, paving the way for next-generation AO.

## ACKNOWLEDGMENTS

Marcos van Dam of Flat Wavefronts provided useful conversations regarding AO error budgets. Brian Bauman and Dave Palmer of LLNL and Chris Lockwood, Ellie Gates, John Gates, Bob Kibrick, and Jim Ward of Lick Observatory have provided support to the ViLLaGEs project over the years. Scott Severson, formerly of the Center for Adaptive Optics and currently of Sonoma State University, co-pioneered ViLLaGEs.

This work was performed under the Michelson Graduate Fellowship for the Jet Propulsion Laboratory, California Institute of Technology, sponsored by the United States Government under a Prime Contract between the California Institute of Technology and NASA. The Villages experiment was supported through a Small Grant for Exploratory Research from the Astronomy Division of the National Science Foundation, award number 0649261. This research was supported in part by the University of California and National Science Foundation Science and Technology Center for Adaptive Optics, managed by the University of California at Santa Cruz under cooperative agreement No. AST-9876783, and directed by Claire E. Max. Support for this work was also provided by a grant from the Gordon and Betty Moore Foundation to the Regents of the University of California, Santa Cruz, on behalf of the UCO/Lick Laboratory for Adaptive Optics. Portions of this work were performed under the auspices of the U. S. Department of Energy by the University of California, Lawrence Livermore National Laboratory under Contract W-7405-ENG-48.

This research has made use of the SIMBAD database and NASA's Astrophysics Data System.